\PassOptionsToPackage{unicode}{hyperref}
\PassOptionsToPackage{hyphens}{url}
\documentclass[
]{article}
\usepackage{amsmath,amssymb}
\usepackage{lmodern}
\usepackage{iftex}
\ifPDFTeX
  \usepackage[T1]{fontenc}
  \usepackage[utf8]{inputenc}
  \usepackage{textcomp} 
\else 
  \usepackage{unicode-math}
  \defaultfontfeatures{Scale=MatchLowercase}
  \defaultfontfeatures[\rmfamily]{Ligatures=TeX,Scale=1}
\fi
\IfFileExists{upquote.sty}{\usepackage{upquote}}{}
\IfFileExists{microtype.sty}{
  \usepackage[]{microtype}
  \UseMicrotypeSet[protrusion]{basicmath} 
}{}
\makeatletter
\@ifundefined{KOMAClassName}{
  \IfFileExists{parskip.sty}{%
    \usepackage{parskip}
  }{
    \setlength{\parindent}{0pt}
    \setlength{\parskip}{6pt plus 2pt minus 1pt}}
}{
  \KOMAoptions{parskip=half}}
\makeatother
\usepackage{xcolor}
\usepackage[margin=1in]{geometry}
\usepackage{graphicx}
\makeatletter
\def\maxwidth{\ifdim\Gin@nat@width>\linewidth\linewidth\else\Gin@nat@width\fi}
\def\maxheight{\ifdim\Gin@nat@height>\textheight\textheight\else\Gin@nat@height\fi}
\makeatother
\setkeys{Gin}{width=\maxwidth,height=\maxheight,keepaspectratio}
\makeatletter
\def\fps@figure{htbp}
\makeatother
\renewcommand{\and}{\\}
\usepackage{booktabs}
\usepackage{longtable}
\usepackage{array}
\usepackage{multirow}
\usepackage{wrapfig}
\usepackage{float}
\usepackage{colortbl}
\usepackage{pdflscape}
\usepackage{tabu}
\usepackage{threeparttable}
\usepackage{threeparttablex}
\usepackage[normalem]{ulem}
\usepackage{makecell}
\usepackage{xcolor}
\ifLuaTeX
  \usepackage{selnolig}  
\fi
\IfFileExists{bookmark.sty}{\usepackage{bookmark}}{\usepackage{hyperref}}
\IfFileExists{xurl.sty}{\usepackage{xurl}}{} 
\hypersetup{
  pdftitle={A Statistical Inquiry into Gender-Based Income Inequality in Canada},
  pdfauthor={Ali R. Kaazempur-Mofrad; Faculty of Arts \& Sciences, University of Toronto},
  hidelinks,
  pdfcreator={LaTeX via pandoc}}

\title{A Statistical Inquiry into Gender-Based Income Inequality in
Canada}
\author{Ali R. Kaazempur-Mofrad\footnote{Currently\(:\) Department of
  Statistics, University of California, Los Angeles (Email\(:\)
  \href{mailto:amofrad@ucla.edu}{\nolinkurl{amofrad@ucla.edu}})} \and Faculty
of Arts \& Sciences, University of Toronto}
\date{}

\begin{document}
\maketitle

\hypertarget{abstract}{%
\section{Abstract}\label{abstract}}

Income inequality distribution between social groups has been a global
challenge. The focus of this study is to investigate the potential
impact of female income on family size and purchasing power. Using
statistical methods such as simple linear regression, maximum likelihood
analysis, and hypothesis testing, I evaluated and investigated the
variability of female pre-tax income with respect to family size. The
results obtained from this study illustrate that for each additional
household member, the average purchasing power decreases. Additionally,
the Bayesian analysis indicates that the probability for an individual
with a pre-tax income of at least one and two standard deviations above
the population mean is female is approximately 1/3 and ¼, respectively,
further highlighting the gender-based income inequality in Canada. This
analysis concludes that although female pre-tax income has no
statistically significant impact on family size, the female pre-tax
income per person has a statistically significant impact on family size.

\hypertarget{introduction}{%
\section{1. Introduction}\label{introduction}}

Inequality in any society is due to many social, economic and political
factors. Income distribution amongst different groups within a society
is one of these underlying issues. Unequal income distribution results
in challenges faced by different groups within the population. Women, in
particular, are faced with income inequality that causes various social
and economic challenges for them and could affect their decision making
and lifestyle. This study focuses on exploring the impact of female
income on family size.

The underlying question that this analysis aims to address is whether
female income impacts family size and purchase power. Specifically, the
objective of this study is to understand whether and how the total, and
per person, female income relates to family size. For the purpose of
this study, it is hypothesized that female income presents an impact on
family size. To evaluate and test this hypothesis I conduct statistical
methods and approaches as presented in the next sections. To further
inform this analysis and to highlight income inequality, I also
investigate the probability that higher income earners are female.

\hypertarget{terminology}{%
\subsubsection{1.1. Terminology}\label{terminology}}

The terms used in this study are introduced below:

\begin{itemize}
\item
  Family -- For the purpose of this analysis, a family contains a
  minimum of 2 individuals living in the same household
\item
  Family size - The number of household members in a family
\item
  Household size - Family size for each household
\item
  Income -- The income used in this study is the pre-tax income
\item
  Pre-tax income - Gross earnings prior to taxation
\item
  Pre-tax income per person - The pre-tax income divided by family size
\item
  Purchase power - The financial ability of an individual to purchase
  goods
\end{itemize}

\hypertarget{data}{%
\section{2. Data}\label{data}}

The data used for this analysis was collected from the 2017 Canadian
Income Survey as a sub-sample to the respondents of the Labour Force
Survey. The data from this survey was obtained from ODESI
{[}1{]}\footnote{Ontario Data Documentation, Extraction Service and
  Infrastructure (ODESI) is a ``web-based data exploration, extraction
  and analysis tool for social science data''
  (\url{http://odesi2.scholarsportal.info/webview/})}, a data portal
available to researchers, teachers and students.

\hypertarget{data-cleaning}{%
\subsubsection{2.1. Data Cleaning}\label{data-cleaning}}

As responses in this data were all numerical, in order to clean up the
data for the analysis, I recoded the responses for Sex from `1' and `2'
to `MALE' and `FEMALE'.

Additionally, for those who chose to skip the income responses, the
value entered in the data was 99999996. As such, in order to prevent
these values from skewing the results, I filtered out these responses to
ensure accurate measures.

\hypertarget{important-variables}{%
\subsubsection{2.2. Important Variables}\label{important-variables}}

The variables used in this study are introduced below:

\begin{itemize}
\item
  \texttt{household\_size} -- Family size for each household
\item
  \texttt{pre\_tax} -- Average pre-tax income for each family size
\item
  \texttt{pre\_tax\_sd} -- Standard deviation of pre-tax income for each
  family size.
\item
  \texttt{pre\_tax\_per} -- Average pre-tax income for each person
  within a family
\item
  \texttt{pre\_tax\_per\_sd} -- Standard deviation of pre-tax income for
  each person within a family
\item
  \texttt{n} -- Number of observations
\end{itemize}

The statistical analysis for this study was conducted using the
programming language R.

\hypertarget{numerical-summaries}{%
\subsubsection{2.3. Numerical Summaries}\label{numerical-summaries}}

Summarizing the female pre-tax income values, both total and per person:

\begin{table}[H]
\centering
\begin{tabular}[t]{l|r|r|r|r|r|r}
\hline
SEX & household\_size & pre\_tax & pre\_tax\_sd & pre\_tax\_per & pre\_tax\_per\_sd & n\\
\hline
FEMALE & 2 & 37224.03 & 33871.03 & 18612.015 & 16935.514 & 15136\\
\hline
FEMALE & 3 & 39683.16 & 34775.13 & 13227.720 & 11591.711 & 6309\\
\hline
FEMALE & 4 & 42400.20 & 40474.50 & 10600.049 & 10118.625 & 6011\\
\hline
FEMALE & 5 & 37572.39 & 34901.75 & 7514.477 & 6980.350 & 2037\\
\hline
FEMALE & 6 & 35466.25 & 34037.52 & 5911.042 & 5672.921 & 563\\
\hline
FEMALE & 7 & 35295.74 & 27746.34 & 5042.249 & 3963.763 & 223\\
\hline
\end{tabular}
\end{table}

For the average female pre-tax income per person, it is helpful to view
the data within a range of values with a 95\% confidence interval.

To calculate 95\% confidence interval:

CI: \(\bar{X} \pm Z_{\frac{\alpha}{2}}\frac{\sigma}{\sqrt{n}}\)

\begin{itemize}
\item
  We are 95\% confident that the mean of female pre-tax income per
  person in a household of 2 people is between \$18,342.22 and
  \$18881.81 CAD.
\item
  We are 95\% confident that the mean of female pre-tax income per
  person in a household of 3 people is between \$12,941.69 and
  \$13,513.75 CAD
\item
  We are 95\% confident that the mean of female pre-tax income per
  person in a household of 4 people is between \$10344.25 and \$10855.85
  CAD
\item
  We are 95\% confident that the mean of pre-tax income per person in a
  household of 5 people is between \$7211.35 and \$7817.61 CAD.
\item
  We are 95\% confident that the mean of pre-tax income per person in a
  household of 6 people is between \$5442.44 and \$6379.64 CAD
\item
  We are 95\% confident that the mean of pre-tax income per person in a
  household of 7 people is between \$4522.01 and \$5562.49 CAD
\end{itemize}

In order to visualize the breakdown of average female pre-tax income
with respect to different family sizes, the following plot (Figure 1) is
generated:

\includegraphics{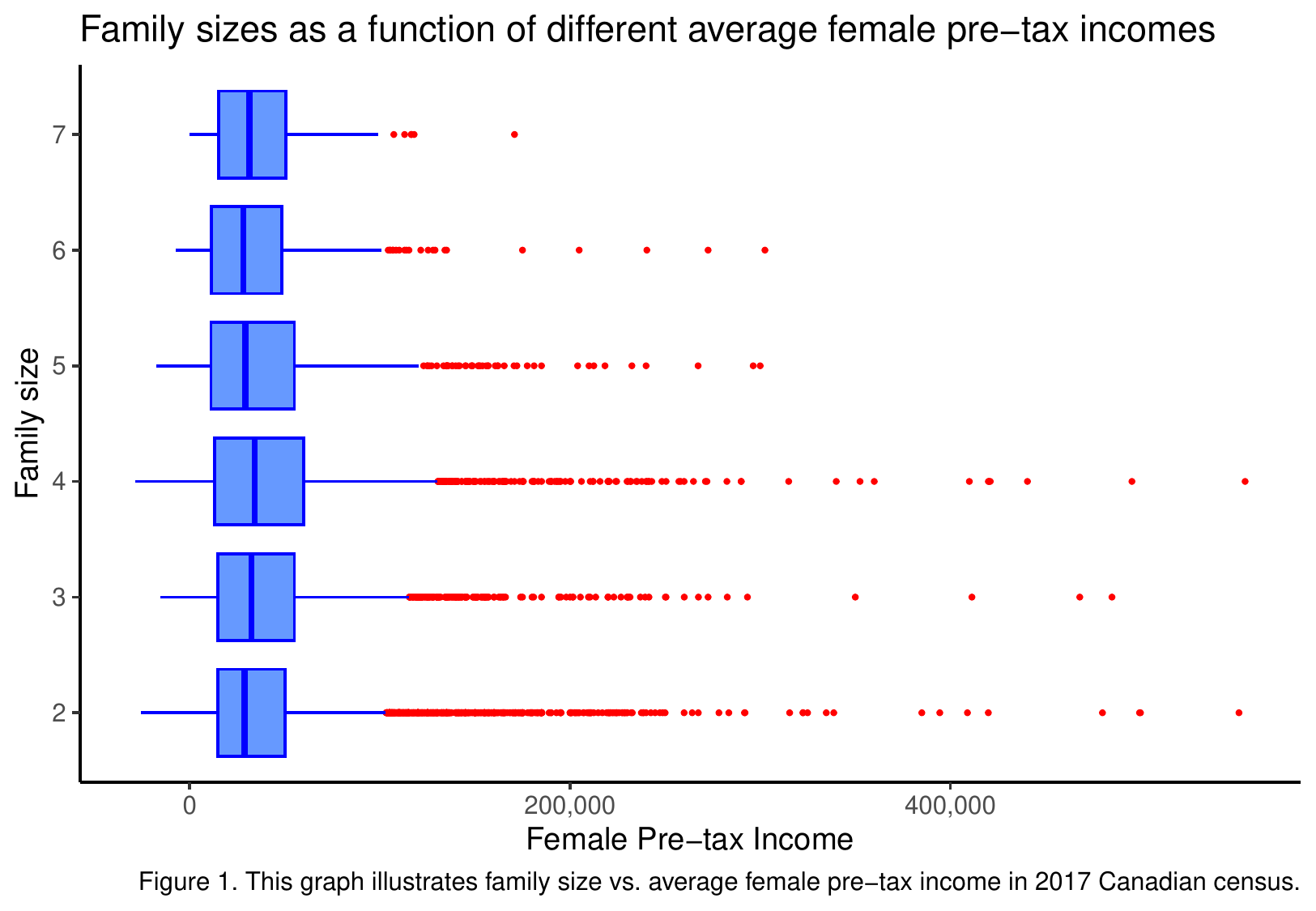}

It is noteworthy that, as evident in Figure 1, individuals can report
negative income values under certain earning sources, eg.
self-employment or investments, which can yield negative income.

\hypertarget{methods}{%
\section{3. Methods}\label{methods}}

For this study, I assumed that a family consists of households that
include a minimum of 2 individuals. This assumption is consistent with
the Census definition of a family. As such, I have not included single
person households in this analysis.

The primary parameters of interest used in this study are the average
and standard deviation of female total, and per person, pre-tax incomes
as well as household size.

I've used the following statistical methodologies to analyze the data:

\textbf{Simple linear regression} -- This model illustrates the rate of
change of female income as the household size increases

\textbf{Confidence interval} -- This methodology allows us to provide a
range of values in which we are 95\% confident that the mean of female
pre-tax income per person lies between

\textbf{Maximum likelihood estimation} -- This frequentist model allows
us to estimate parameters that fit the female pre-tax income data.
Through exploratory data analysis (EDA), it is evident that the spread
of female pre-tax incomes represent a right skewed normal distribution,
hence it is justified to use a normal distribution for this model.

\textbf{Hypothesis test} -- This methodology allows us to determine the
probability in which the difference of two sample means is at least as
large as observed, under the assumption of the null hypothesis

\textbf{Bayesian credible interval} -- To inform my analysis and
illustrate the gender-based income inequalities in Canada, I conducted a
Bayesian modelling analysis. This model allows us to determine the
probability in which an individual with a pre-tax income of at least
\$100,000 CAD is a female earner. I initially chose this income level as
a symbolic value, but later I added two more tests where I implemented
values of 1 and 2 standard deviations above the population's (\emph{male
and female}) mean individual pre-tax income. As such, the two additional
Bayesian models were created using \$95,314.06 \footnote{Mean of
  population pre-tax income with the addition of 1 standard deviation}
and \$143,235.50 \footnote{Mean of population pre-tax income with the
  addition of 2 standard deviations} CAD. In other words, these two
latter models illustrate the probability in which an individual with a
pre-tax income of at least 1 and 2 standard deviations above the mean of
the population's (male and female) mean individual pre-tax income is a
female earner.

The prior for this model was chosen based upon the ratio of females and
males in the study to the total participants who answered the question
of pre-tax income in the survey. The likelihood for this model was
chosen based on the ratio of females and males in the study who earned a
pre-tax income of at least \$100,000 CAD (\$95,314.06 and \$143,235.50
CAD for the other models).

\hypertarget{results}{%
\section{4. Results}\label{results}}

In this section, I present models, results and their interpretations to
describe the analysis of the female pre-tax income data vis-à-vis
household size.

\hypertarget{regression-analysis}{%
\subsubsection{4.1. Regression Analysis}\label{regression-analysis}}

To better understand the trends in the data, the average female pre-tax
income per person with respect to different family sizes is visualized
using a \textbf{simple linear regression model} (Figure 2).
Specifically, this model is used to estimate the rate of change of
income per person with an increase of family size.

\includegraphics{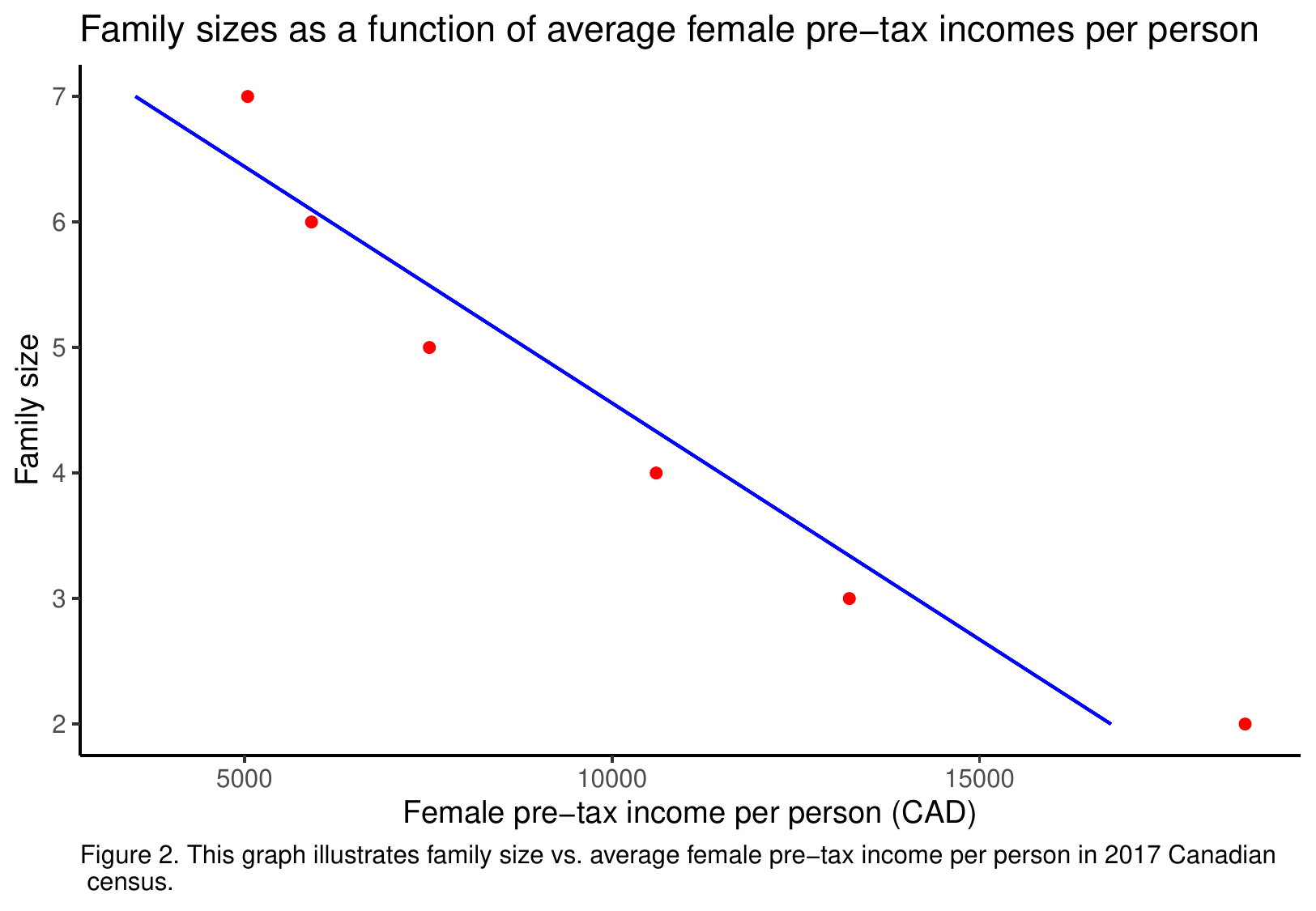}

From the simple linear regression model, it was determined that for each
additional household member, on average, the pre-tax female income per
person decreased by \$2653.8 CAD. This illustrates that as family size
increases, the average income allotted to each household member
decreases and thus limits each individual's purchasing power (see Figure
2). This observation seems reasonable as the total income doesn't change
considerably (see Figure 1 and also see hypothesis test results below)
while the overall household size increases.

\hypertarget{maximum-likelihood-estimation-analysis}{%
\subsubsection{4.2. Maximum Likelihood Estimation
Analysis}\label{maximum-likelihood-estimation-analysis}}

In this section, I examine the spread of total female pre-tax incomes,
which allows for a better understanding of the mean and median values as
well as the behavior of the data's spread (Figure 3).
\includegraphics{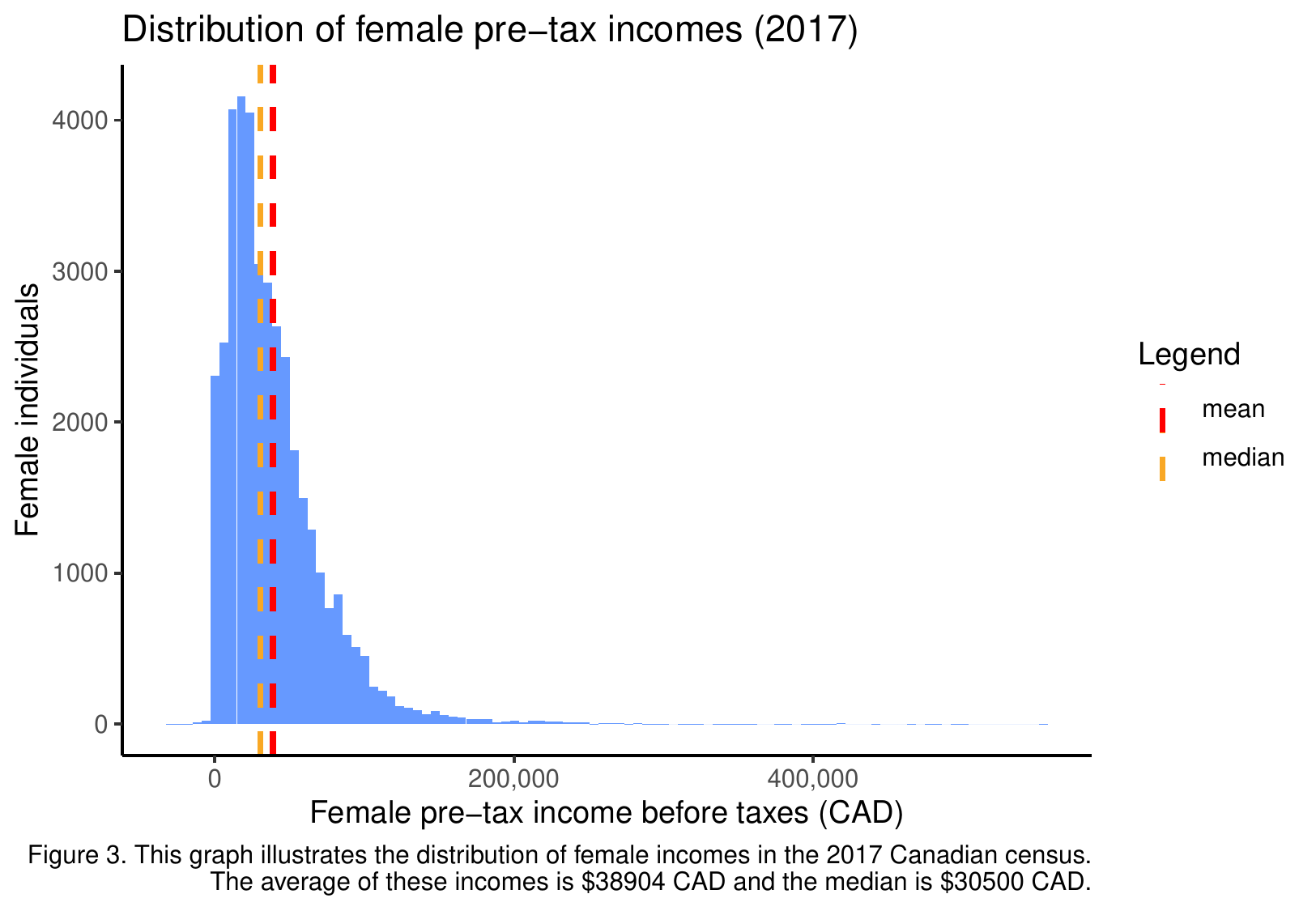}

To fit a model for the analysis of the total pre-tax income for women, a
\emph{univariate frequentist} model can be used. The data collected is
used as sample data to form a \emph{maximum likelihood estimation}
model. We will use a \emph{normal distribution} because using
exploratory data analysis (EDA), it is evident that the data is spread
across 2 tails with a right skewed normal distribution. The model
estimation used for this purpose is:

\[Y_{i,j} \sim N(\mu_j, \sigma^2_i)  \] where \(Y_{i,j}\) is the pre-tax
income for female i with household size j. \(\mu_j\) is the average
female pre-tax income for household size j and \(\sigma^2_i\) is the
variance for household j.

Using the \emph{maximum likelihood estimation model}, the parameter
estimations for the mean and variance of the normally distributed female
pre-tax incomes are obtained.

The \(\hat{\mu}_{MLE}\) estimates for each household size are presented
in the following table:

\begin{table}[H]
\centering
\begin{tabular}[t]{r|r}
\hline
household\_size & mu\_MLE (CAD)\\
\hline
2 & 37224.03\\
\hline
3 & 39683.16\\
\hline
4 & 42400.20\\
\hline
5 & 37572.39\\
\hline
6 & 35466.25\\
\hline
7 & 35295.74\\
\hline
\end{tabular}
\end{table}

Likewise, the \(\hat{\sigma}^2_{MLE}\) estimates for each household size
are presented in the following table:

\begin{table}[H]
\centering
\begin{tabular}[t]{r|r}
\hline
household\_size & Sigma\_Squared\_MLE\\
\hline
2 & 1147170790\\
\hline
3 & 1209118273\\
\hline
4 & 1637912729\\
\hline
5 & 1217534310\\
\hline
6 & 1156495193\\
\hline
7 & 766407336\\
\hline
\end{tabular}
\end{table}

\hypertarget{hypothesis-testing}{%
\subsubsection{4.3. Hypothesis Testing}\label{hypothesis-testing}}

In this section, I present the hypotheses I proposed in relation to my
underlying question of whether female income impacts family size and
purchase power. Originally, I hypothesized that female pre-tax income
impacts family size (\emph{Hypothesis I}). However, as described below
in my hypothesis testing, this hypothesis is rejected. Then, I tested a
second hypothesis that female pre-tax income per person impacts family
size (\emph{Hypothesis II}). This hypothesis failed to be rejected and
showed a statistically significant change in the mean female pre-tax
income per person in a household.

In the next two subsections, I present the details of my hypothesis
testing.

\hypertarget{testing-hypothesis-i}{%
\subsubsection{\texorpdfstring{4.3.1. Testing \emph{Hypothesis
I}}{4.3.1. Testing Hypothesis I}}\label{testing-hypothesis-i}}

In this subsection, I conduct hypothesis testing for Hypothesis I,
proposing that female pre-tax income impacts family size. Specifically,
to conduct this test, I compare the mean values of female pre-tax income
for family sizes of 2 vs.~7, 2 vs.~5, and 5 vs.~7 as presented below,
respectively.

\[H_0:\ \mu_{female\ income,\  2\ people} = \mu_{female\ income,\  7\ people}\]
\[H_a:\ \mu_{income\ per\ person,\ 2\ people} \ne \mu_{income\ per\ person,\ 7\ people}\]

\[T=\frac{\bar{x}_2-\bar{x}_1}{\sqrt{\frac{s_1^2}{n_1}+\frac{s_2^2}{n_2}}}\]

\[=\frac{35295.74-37224.03}{\sqrt{\frac{33871.03^2}{15136}+\frac{27746.34^2}{223}}}=-1.026605\]

Through the R function \texttt{pt()}, we calculate the p-value to be
0.152. As the p-value is greater than the 0.05 threshold, we fail to
reject the null. This suggests that there is a 15.2\% percent chance
that the difference between the sample means of female pre-tax income
for family sizes of 2 and 7 is at least as large as observed, under the
assumption of the null hypothesis where the means are equal. With this
value being above the 0.05 threshold, we fail to reject the hypothesis
that the mean incomes are equal. This does not however allow us to
``accept'' the alternate hypothesis of the mean values being unequal. In
lay terms, the female pre-tax income is not different for family sizes
of 2 and 7 in a statistically significant manner.

\[H_0:\ \mu_{female\ income,\ 2\ people} = \mu_{female\ income,\ 5\ people}\]
\[H_a:\ \mu_{female\ income,\ 2\ people} \ne \mu_{female\ income,\ 5\ people}\]

\[T=\frac{\bar{x}_2-\bar{x}_1}{\sqrt{\frac{s_1^2}{n_1}+\frac{s_2^2}{n_2}}}\]

\[=\frac{37572.39-37224.03}{\sqrt{\frac{33871.03^2}{15136}+\frac{34901.75^2}{2037}}}=0.4243881\]

Through the R function \texttt{pt()}, we calculate the p-value to be
0.336. As the p-value is greater than the 0.05 threshold, we fail to
reject the null. This suggests that there is a 33.6\% percent chance
that the difference between the sample means of female pre-tax income
for family sizes of 2 and 5 is at least as large as observed, under the
assumption of the null hypothesis where the means are equal. With this
value being above the 0.05 threshold, we fail to reject the hypothesis
that the mean incomes are equal. This does not however allow us to
``accept'' the alternate hypothesis of the mean values being unequal. In
lay terms, the female pre-tax income is not different for family sizes
of 2 and 5 in a statistically significant manner.

\[H_0:\ \mu_{female\ income,\ 5\ people} = \mu_{female\ income,\ 7\ people}\]
\[H_a:\ \mu_{female\ income,\ 5\ people} \ne \mu_{female\ income,\ 7\ people}\]

\[T=\frac{\bar{x}_2-\bar{x}_1}{\sqrt{\frac{s_1^2}{n_1}+\frac{s_2^2}{n_2}}}\]

\[=\frac{35295.74-37572.39}{\sqrt{\frac{34901.75^2}{2037}+\frac{27746.34^2}{223}}}=-1.131237\]

Through the R function \texttt{pt()}, we calculate the p-value to be
0.129. As the p-value is greater than the 0.05 threshold, we fail to
reject the null. This suggests that there is a 12.9\% percent chance
that the difference between the sample means of female pre-tax income
for family sizes of 5 and 7 is at least as large as observed, under the
assumption of the null hypothesis where the means are equal. With this
value being above the 0.05 threshold, we fail to reject the hypothesis
that the mean incomes are equal. This does not however allow us to
``accept'' the alternate hypothesis of the mean values being unequal. In
lay terms, the female pre-tax income is not different for family sizes
of 5 and 7 in a statistically significant manner.

From the results of the above hypothesis tests, Hypothesis I, stating
that female pre-tax income impacts family size, is rejected. In the next
subsection, I conduct hypothesis testing for Hypothesis II.

\hypertarget{testing-hypothesis-ii}{%
\subsubsection{\texorpdfstring{4.3.2. Testing \emph{Hypothesis
II}}{4.3.2. Testing Hypothesis II}}\label{testing-hypothesis-ii}}

In this subsection, I conduct hypothesis testing for Hypothesis II,
proposing that female pre-tax income impacts family size. Specifically,
to conduct this test, I compare the mean values of female pre-tax income
per person for family sizes of 2 vs.~7, 5 vs.~7, and 2 vs.~5 as
presented below, respectively.

\[H_0:\ \mu_{female\ income\ per\ person,\ 2\ people} = \mu_{female\ income\ per\ person,\ 7\ people}\]
\[H_a:\ \mu_{female\ income\ per\ person,\ 2\ people} \ne \mu_{female\ income\ per\ person,\ 7\ people}\]

\[T=\frac{\bar{x}_2-\bar{x}_1}{\sqrt{\frac{s_1^2}{n_1}+\frac{s_2^2}{n_2}}}\]
\[=\frac{5042.249-18612.015}{\sqrt{\frac{3963.763^2}{223}+\frac{16935.514^2}{15136}}}=-45.38315\]

Through the R function \texttt{pt()}, we calculate the p-value to be 0.
As the p-value is less than the 0.05 threshold, we reject the null. This
suggests that there is a 0\% chance that the difference between the
sample means of income per person for family sizes of 2 and 7 is at
least as large as observed, under the assumption of the null hypothesis
where the means are equal. With this value being below the 0.05
threshold, we have statistically significant evidence to reject the null
hypothesis that the mean incomes per person are equal. This tells us
that there is a significant difference between the average income per
person for family sizes of 2 and 7.

\[H_0:\ \mu_{female\ income\ per\ person,\ 5\ people} = \mu_{female\ income\ per\ person,\ 7\ people}\]
\[H_a:\ \mu_{female\ income\ per\ person,\ 5\ people} \ne \mu_{female\ income\ per\ person,\ 7\ people}\]

\[T=\frac{\bar{x}_2-\bar{x}_1}{\sqrt{\frac{s_1^2}{n_1}+\frac{s_2^2}{n_2}}}\]
\[=\frac{5042.249-7514.477}{\sqrt{\frac{3963.763^2}{223}+\frac{6980.350^2}{2037}}}=-8.047486\]

Through the R function \texttt{pt()}, we calculate the p-value to be
\(7.105*10^{-16}\). As the p-value is less than the 0.05 threshold, we
reject the null. This suggests that there is a \(7.105*10^{-14}\)\%
chance that the difference between the sample means of income per person
for family sizes of 2 and 5 is at least as large as observed, under the
assumption of the null hypothesis where the means are equal. With this
value below the 0.05 threshold, we have statistically significant
evidence to reject the null hypothesis that the mean incomes per person
are equal. This tells us that there is a significant difference between
the average income per person for family sizes of 2 and 5.

\[H_0:\ \mu_{female\ income\ per\ person,\ 2\ people} = \mu_{female\ income\ per\ person,\ 5\ people}\]
\[H_a:\ \mu_{female\ income\ per\ person,\ 2\ people} \ne \mu_{female\ income\ per\ person,\ 5\ people}\]

\[T=\frac{\bar{x}_2-\bar{x}_1}{\sqrt{\frac{s_1^2}{n_1}+\frac{s_2^2}{n_2}}}\]
\[=\frac{7514.477-18612.015}{\sqrt{\frac{16935.514^2}{15136}+\frac{6980.350^2}{2037}}}=-53.59875\]

Through the R function \texttt{pt()}, we calculate the p-value to be 0.
As the p-value is less than the 0.05 threshold, we reject the null. This
suggests that there is a 0\% chance that the difference between the
sample means of income per person for family sizes of 2 and 5 is at
least as large as observed, under the assumption of the null hypothesis
where the means are equal. With this value below the 0.05 threshold, we
have statistically significant evidence to reject the null hypothesis
that the mean incomes per person are equal. This tells us that there is
a significant difference between the average income per person for
family sizes of 2 and 5.

In summary, from the above hypothesis tests, we fail to reject
Hypothesis II, stating that female pre-tax income per person impacts
family size in a statistically significant manner.

\hypertarget{bayesian-analysis}{%
\subsubsection{4.4. Bayesian Analysis}\label{bayesian-analysis}}

In this section, I conduct Bayesian analysis which involves obtaining
posterior probabilities through the use of prior and likelihood
probabilities. Specifically, I use Bayesian analysis to estimate
\emph{the probability of an individual who earns above a certain
threshold of pre-tax income to be female}. For this threshold value of
pre-tax income, I use three different values. First, I use \$100,000 CAD
as a symbolic 6-figure income threshold. Then, to be more technical, I
use 1 and 2 standard deviations above the mean of population\footnote{Note:
  This is the mean of all individual (male and female) pre-tax incomes}
pre-tax income.

In this first model, I use Bayesian analysis to obtain the probability
in which an individual who earns a pre-tax income of at least \$100,000
CAD is a female earner.

F = event that case is a female

W = event that income is greater than \$100,000 CAD

Prior probabilities: \[P(F) = \frac{38585}{75372} = 0.5119275\]

\[P(F^c) = 1-0.5119275 = 0.4880725\]

Likelihood probabilities:

\[P(W | F) = \frac{1873}{75372} = 0.02485008\]
\[P(W | F^c) = \frac{4880}{75372} = 0.06474553\]

\[P(F\ |\ W)=\frac{P(F\  \cap\  W)}{P(W)}  = \frac{P(W\ |\ F)*P(F)}{P(W\ |\ F)*P(F) + P(W\ |\ F^c)*P(F^c)}\]

\[=\frac{0.02485008*0.5119275}{0.02485008*0.5119275+0.06474553*0.4880725} = 0.2870235\]
\[P(F\ |\ W)=0.287\]

This implies that there is a \textbf{28.7\%} chance that an individual
who earns a pre-tax income of at least \$100,000 CAD is a female.

Creating another Bayesian model to obtain the probability in which an
individual who earns a pre-tax income of at least \$95,314.06 CAD
\footnote{Mean of population pre-tax income with the addition 1 standard
  deviation} is female:

F = event that case is a female

W = event that income is greater than \$95,314.06 CAD

Prior probabilities: \[P(F) = 36787/75372 = 0.5119275\]

\[P(F^c) = 1-0.4880725 = 0.4880725\]

Likelihood probabilities:

\[P(W | F) = 2182/75372 = 0.02894974\]
\[P(W | F^c) = 5317/75372 = 0.07054344\]

\[P(F\ |\ W)=\frac{P(F\  \cap\  W)}{P(W)}  = \frac{P(W\ |\ F)*P(F)}{P(W\ |\ F)*P(F) + P(W\ |\ F^c)*P(F^c)}\]

\[=\frac{0.02894974*0.5119275}{0.02894974*0.5119275+0.07054344*0.4880725} = 0.3009142\]
\[P(F\ |\ W)=0.301\]

This implies that there is a \textbf{30.1\%} chance that an individual
who earns a pre-tax income of at least \$95,314.06 CAD is a female.

Creating another Bayesian model to obtain the probability in which an
individual who earns a pre-tax income of at least \$143,235.50 CAD
\footnote{Mean of population pre-tax income with the addition 2 standard
  deviations} is female:

F = event that case is a female

W = event that income is greater than \$143,235.50 CAD

Prior probabilities: \[P(F) = 36787/75372 = 0.5119275\]

\[P(F^c) = 1-0.4880725 = 0.4880725\]

Likelihood probabilities:

\[P(W | F) = 581/75372 = 0.007708433\]
\[P(W | F^c) = 1816/75372 = 0.02409383\]

\[P(F\ |\ W)=\frac{P(F\  \cap\  W)}{P(W)}  = \frac{P(W\ |\ F)*P(F)}{P(W\ |\ F)*P(F) + P(W\ |\ F^c)*P(F^c)}\]

\[=\frac{0.007708433*0.5119275}{0.007708433*0.5119275+0.02409383*0.4880725} = 0.2512566\]
\[P(F\ |\ W)=0.251\]

This implies that there is a \textbf{25.1\%} chance that an individual
who earns a pre-tax income of at least \$143,235.50 CAD is a female.

In summary, the above Bayesian analyses suggest that the probability for
an individual who earns at least \$100,000 CAD (a symbolic 6-figure
income threshold) is 28.7\%. More technically relevant, the probability
that an individual who earns a pre-tax income 1 and 2 standard
deviations above the population pre-tax individual mean is a female is
30.1\% and 25.1\%, respectively. This clearly reflects the gender-based
income inequalities in Canada highlighting that as the income bracket
increases, the ratio of females to males appears to decrease.

\hypertarget{conclusion}{%
\section{5. Conclusion}\label{conclusion}}

This study explored the question of whether female pre-tax income
impacts household size. The original hypothesis that drove this
investigation proposed that female income impacts family size. To
address the question and test the hypothesis, I used statistical methods
including simple linear regression, maximum likelihood estimation, and
hypothesis testing. To further inform my analysis, I also conducted a
Bayesian analysis to highlight the income inequality that females
experience in Canada.

This analysis generated some interesting results that showed that female
pre-tax income has no statistically significant impact on family size.
This result leads to a rejection of my original hypothesis. When I
examined female pre-tax income normalized by family size (i.e., female
income per person), the results illustrated that for each additional
household member, the average income per person decreased linearly in a
statistically significant manner. This study concludes that female
pre-tax income has no statistically significant impact on family size,
whereas the female pre-tax income per person has a statistically
significant impact on family size.

In terms of limitations, the scope of methodologies and models used for
this inquiry do not entirely allow for establishing correlation and
deeper understanding and analysis of the subject. Another potential
limitation for this analysis could be that the survey participants be
skewed towards people with higher education and possibly higher income.
Therefore, this survey may not accurately reflect the entire population
of Canada.

To further this investigation, more analysis needs be conducted to
explore whether and how tax policies might affect the results of female
income on family size.

\hypertarget{bibliography}{%
\section{Bibliography}\label{bibliography}}

{[}1{]} Statistics Canada. (2018). Canadian Income Survey, 2017
{[}Canadian Income Survey{]}. Income Statistics Division {[}Producer{]}.
Data Liberation Initiative {[}Distributor{]}.
\url{http://odesi2.scholarsportal.info/webview/index.jsp?v=2\&submode=abstract\&study=http\%3A\%2F\%2F142.150.190.128\%3A80\%2Fobj\%2FfStudy\%2FCIS-72M0003-E-2017\&mode=documentation\&top=yes}

\hypertarget{appendix}{%
\section{Appendix}\label{appendix}}

\hypertarget{derivation-of-maximum-likelihood-estimator}{%
\subsubsection{Derivation of Maximum Likelihood
Estimator:}\label{derivation-of-maximum-likelihood-estimator}}

PDF of Normal Distribution:
\(f(y_i\  |\  \mu,\sigma^2) =\frac{1}{\sqrt{2\pi\sigma^2}}*exp({-\frac{1}{2}}(\frac{y_i - \mu}{\sigma})^2)\)

Likelihood Function:
\[L(\mu,\sigma^2) = \prod_{i=1}^{n} f(y_i\  |\  \mu,\sigma^2)\]
\[= \prod_{i=1}^{n}\frac{1}{\sqrt{2\pi\sigma^2}}*exp({-\frac{1}{2}}(\frac{y_i - \mu}{\sigma})^2)\]
\[=(2\pi\sigma^2)^{-\frac{n}{2}}exp[{\sum_{i=1}^{n}-\frac{1}{2}(\frac{y_i-\mu}{\sigma})^2}]\]

\[=(2\pi\sigma^2)^{-\frac{n}{2}}exp[-\frac{1}{2\sigma^2}{\sum_{i=1}^{n}(y_i-\mu)^2}]\]

Log-likelihood Function:
\[l(\mu,\sigma^2)=-\frac{n}{2}log(2\pi\sigma^2)-\frac{1}{2\sigma^2}\sum_{i=1}^{n}(y_i-\mu)^2\]

\[=-\frac{n}{2}log(2\pi)-\frac{n}{2}log(\sigma^2)-\frac{1}{2\sigma^2}\sum_{i=1}^{n}(y_i-\mu)^2\]

\hypertarget{partial-derivatives}{%
\subsection{Partial derivatives:}\label{partial-derivatives}}

For \(\hat{\mu}_{MLE}:\)

\[\frac{\delta l(\mu,\sigma^2)}{\delta \mu} = -\frac{2}{2\sigma^2}\sum_{i=1}^{n}(y_i-\mu)(-1) = \frac{1}{\sigma^2}[\sum_{i=1}^{n}(y_i) - n\mu]\]
Set equal to zero:

\[\frac{1}{\sigma^2}[\sum_{i=1}^{n}(y_i) - n\mu] = 0\]
\[\rightarrow \sum_{i=1}^{n}y_i = n\mu\]
\[\rightarrow \hat{\mu}_{MLE} = \frac{\sum_{i=1}^{n}y_i}{n}\]
\[ \hat{\mu}_{MLE} = \bar{y}\]

For \(\hat{\sigma^2}_{MLE}:\)
\[\frac{\delta l(\mu,\sigma^2)}{\delta \sigma^2} = -\frac{n}{2\sigma^2}+\frac{1}{2(\sigma^2)^2}\sum_{i=1}^{n}(y_i-\mu)^2\]

Set equal to zero:

\[-\frac{n}{2\sigma^2}+\frac{1}{2(\sigma^2)^2}\sum_{i=1}^{n}(y_i-\mu)^2 = 0\]
\[\rightarrow -n + \frac{1}{\sigma^2}\sum_{i=1}^{n}(y_i-\mu)^2=0\]
\[\rightarrow \frac{1}{\sigma^2}\sum_{i=1}^{n}(y_i-\mu)^2 = n\]

\[\hat{\sigma}^2 = \frac{1}{n}\sum_{i=1}^{n}(y_i-\mu)^2\]

Now check \(2^{nd}\) partial derivative:

\[\frac{\delta^2 l(\mu,\sigma^2)}{\delta \mu^2} = \frac{\delta}{\delta \mu}[\frac{1}{\sigma^2}\sum_{i=1}^{n}(y_i-\mu)] = \frac{1}{\sigma^2}(-n) < 0\]

\[\frac{\delta^2 l(\mu,\sigma^2)}{\delta (\sigma^2)^2} = \frac{n}{2(\sigma^2)^2}+\frac{-2}{2(\sigma^2)^3}\sum_{i=1}^{n}(y_i-\mu)^2\]

\[=\frac{n\sigma^2-2\sum_{i=1}^{n}(y_i-\mu)^2}{2(\sigma^2)^3}\] This is
negative for \(n\sigma^2 < 2\sum_{i=1}^{n}(y_i-\mu)^2\)

\[\therefore  \hat{\mu}_{MLE} = \bar{y}\]
\[\hat{\sigma}^2_{MLE} =\frac{1}{n}\sum_{i=1}^{n}(y_i-\mu)^2 \]

\end{document}